# AstroCloud: A Distributed Cloud Computing and Application Platform for Astronomy

Chen-Zhou Cui, Bo-Liang He, Chang-Hua Li, Dong-Wei Fan, Shan-Shan Li, Lin-Ying Mi, Zi-Huang Cao, Si-Si Yang, Yun-Fei Xu, Yue Chen, Zheng Li, Xu Han
Center for Information and Computing ,National Astronomical Observatories, CAS,
Beijing, China
Email: ccz@bao.ac.cn

Ce Yu, Jian Xiao, Zhi Hong, Shucheng Yin, Chen Li
School of Computer Science ,Tianjin University
Tianjin, China

Chuanjun Wang, Yufeng Fan, Jianguo Wang, Junyi Chen
Gaomeigu Observatory ,Yunnan Astronomical Observatories, CAS
Kunming, China

Hailong Zhang
Information Center ,Xinjiang Astronomical Observatory, CAS
Urumqi, China

Liang Liu, Na Gao, Zherui Yang
Information Center, Purple Mountain Observatory, CAS
Nanjing, China

Xiao Chen, Min Liu
Computing Center ,Shanghai Astronomical Observatory, CAS
Shanghai, China

Cuilan Qiao, Kangyu Du
College of Physical Science and Technology ,Center Normal University
Wuhan, China

Liying Su, Wenming Song
College of Mechanical Engineering and Applied Electronics Technology ,Beijing University of Technology
Beijing, China

*Abstract*-Virtual Observatory (VO) is a data-intensively online astronomical research and education environment, which takes advantages of advanced information technologies to achieve seamless and global access to astronomical information. AstroCloud is a cyber-infrastructure for astronomy research initiated by Chinese Virtual Observatory (China-VO) project, and also a kind of physical distributed platform which integrates lots of tasks such as telescope access proposal management, data archiving, data quality control, data release and open access, cloud based data processing and analysis. It consists of five application channels, i.e. observation, data, tools, cloud and public and is acting as a full lifecycle management system and gateway for astronomical data and telescopes. Physically, the platform is hosted in six cities currently, i.e. Beijing, Nanjing, Shanghai, Kunming, Lijiang and Urumqi, and serving more than 17 thousand users. Achievements from international Virtual Observatories and Cloud Computing are adopted heavily. In the paper, backgrounds of the project, architecture, Cloud Computing environment, key features of the system, current status and future plans are introduced.

## I INTRODUCTION

Astronomy in China has been developing fast in the last decades. Large Sky Area Multi-Object Fiber Spectroscopic Telescope (LAMOST) was completed in 2008 and has started sky survey observation since 2009. The construction of Five-hundred-meter Aperture Spherical radio Telescope (FAST) was completed on September 25, 2016 and is now facing the stage of technical commissioning of the telescope. As the early phase of an ambitious plan of Chinese Antarctic Observatory, three telescopes have already been built at Doom A, Antarctic. Wukong (Dark Matter Particle Explorer), the first Astronomical satellite was launched in 2015. More launches are planned in the coming years. In addition to above key projects, a dozen of astronomical observatories with middle-size or small-size telescopes are being constructed. Locations of these facilities are largely distributed across the whole scope of Chinese mainland and even to Argentina and Antarctic. With the requirements of multi-waveband astronomy and time-domain astronomy, astronomical community is faced with big challenges to harvest and utilize data of these facilities.

Chinese Virtual Observatory (China-VO) is the national VO project in China initiated in 2002 [1]. AstroCloud is a cyber-infrastructure for astronomy research initiated by the China-VO under funding support from NDRC (National Development and Reform commission) and CAS (Chinese Academy of Sciences). Taking advantages of Cloud Computing and achievements of Virtual Observatories, a lot of functions are supported on the platform, such as telescope access proposal management, observation data archiving, data quality control, data release and open access, cloud based data processing and analysis and many applications interesting for the public. Based on the internet and virtualization technologies, research resources including scientific data, storage, computing, software and tools are integrated into the online cyber-infrastructure.

The paper is organized as follows. In section 2, architecture of the distributed application system is introduced. Section 3 focuses on Cloud Computing and application environments. Major channels and key features of the platform are described in Section 4. In the last part of the paper, current status and future plans are summarized.

## II PLATFORM ARCHITECTURE

As illustrated in Fig.1, the cyber-infrastructure consists of several sub-systems, which are Telescope (Observation), Data, Cloud, Computing (HPC), one auxiliary tools system, and two fundamental backend components for data archiving and storage [2]. These independent systems are seamlessly integrated into a comprehensive end-to-end astronomy research environment. By this environment, astronomers can make telescope access proposal submission in the Observation system, manage the observed data through the Data system, and perform data processing and analysis using VM (virtual machine) in the Cloud system. In backend, the observed data from telescope can be automatically transferred to nearest local data center which is a node of the AstroCloud, so users can process their data in an in-situ manner within the platform so as to avoid large volume data transfer.

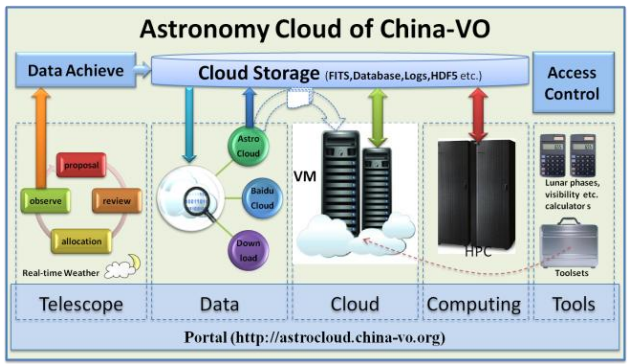

Figure 1. High level overview of the astro cloud platform

The whole platform is built on a distributed cloud storage layer where overall data are resident. In order to reduce the complexity of data management and performance optimization, a distributed data access middleware is designed to provide a unified data access interface. Upper systems can transparently access data without caring the physical location of the data. Moreover, based on the RBAC (role based access control) method and rule engine, a security framework is developed to provide unified user management, guarantee normal data access and protect private data. The platform is mainly implemented by Java programming language, plus a few of C/C++ codes, Python and Bash scripts. As illustrated in Fig. 2, user can have many roles and every role includes many permissions of operation. Each operation is defined by the class and has method and check rule to implement this operation. The rule describes the prerequisite for invoking this operation or the filter criteria for the underlying data. For example, user's request would be mapped to a particular java class and its some method. Security manager will check whether the user has the permission to invoke the class's method, and then if passed, the rule engine will parse and evaluate the rule bound with the operation. According to the type of rule, some filtering constraints may be applied to the following execution for reading operation, or the authorization result will be returned for writing operation.

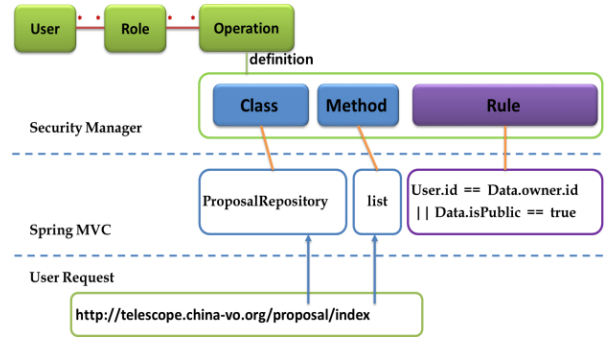

Figure 2. The RBAC and rule based security mechanism

## III CLOUD COMPUTING ENVIRONMENTS

Limited by the poor bandwidth of global internet, it is very difficult and even impossible to download the big astronomical data (TBs to PBs) to user's end. Based on this reason, we purpose that if astronomers have enough computing capabilities in each data center, it will avoid huge data transfer to the local machine. This flow will be more effective. Fig.3 shows two different workflows. The left is the traditional workflow and the right is the new workflow based on AstroCloud.

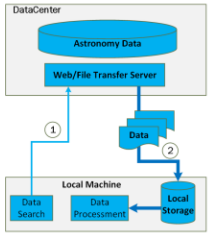 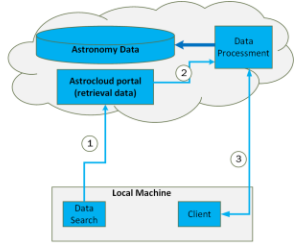

Figure 3. The two different astronomy research workflows

In order to implement this workflow, Cloud Computing technology gives us a perfect solution. Cloud Computing is a model for enabling convenient, on-demand network access to a shared pool of configurable computing resources (e.g., networks, servers, storage, applications, and services) that can be rapidly provisioned and released with minimal management effort or service provider interaction. According to the provisioned resource type, Cloud Computing includes three service models, Software as a Service (SaaS), Platform as a Service (PaaS), and Infrastructure as a Service (IaaS). AstroCloud is a mix of these three service models. [3]

The astronomy data centers are distributed largely in China. At present, there are five professional Astronomical Observatories in China, including National Astronomical Observatories (Beijing), Purple Mountain Observatory (Nanjing), Shanghai Astronomical Observatory (Shanghai), Yunnan Astronomical Observatories (Kunming) and Xinjiang Astronomical Observatory (Urumqi). The AstroCloud platform is currently distributed in six cities covering the five observatories. AstroCloud computing environments integrate these distributed computing, network and storage resources, and give astronomer a centralized resource view and operation platform. Each data center becomes a cloud computing node. Fig.4 shows the architecture of one cloud node.

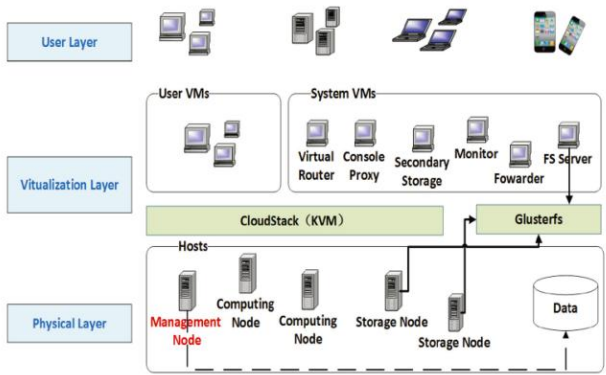

Figure 4. The Architecture of Cloud Computing Environments for AstroCloud

This architecture contains three layers. The top layer is the user layer and provides the user interface including browsers and software clients for many kinds of terminal like desktops or mobiles. The bottom layer is the physical layer to manage different physical resources. The middle layer is the virtualization layer and also is a key content of the AstroCloud platform. CloudStack is used here, which is an open Cloud Computing management middleware software. Based on this platform, we implement the cloud virtual machine creation and access, cloud monitoring, cloud storage, cloud authentication and single sign on. In this Cloud Computing environment, we integrate many IT technologies like VLAN, VPC, Firewall, Port forward, KVM, NFS Samba, etc.

In AstroCloud platform, it includes virtual machine creation and management, cloud storage system of MyVOSpace, cloud monitor and cloud usage log and analysis components. Fig.5 shows the overview of function components and related technology.

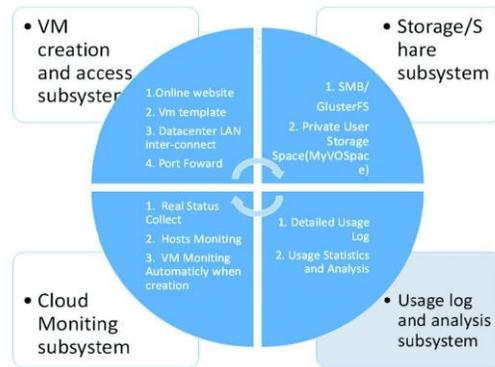

Figure 5. The function components of Cloud Computing Environments

Function 1: Only in few minutes, you can complete a virtual machine (VM) with pre-installed astronomy software. You can access to the created VM by SSH (Secure Shell) or remote desktop from any location. According to your requirement, you can choose the suitable hardware configuration of VM.

Function 2: Provides a private, safe and large capacity storage space, which share and auto-mount to your different VMs.

Function 3: Cloud monitoring system, real-time status collection and notification.

Function 4: Detailed usage log and analysis, including the usage of resource and user billing.

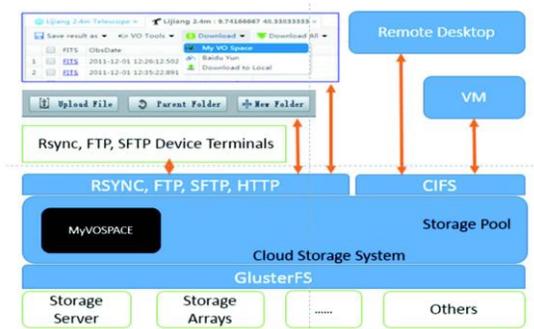

Figure 6. The framework of MyVOSpace of Cloud Computing Environments

MyVOSpace is a private, safe user storage space and also is another key feature of AstroCloud computing environment. In AstroCloud, MyVOSpace can be shared in different VMs and remote desktops under the name of one user. Fig.6 shows the framework of MyVOSpace, cloud storage system and the relation with other components and technologies.

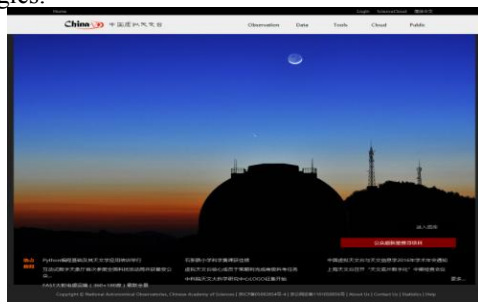

Figure 7. FrontPage of the AstroCloud

IV MAJOR CHANNELS AND KEY FEATURES

AstroCloud was lunched on May 15, 2014. The current system consists of five channels, i.e. Observation, Data, Tools, Cloud and Public, as shown in Fig. 7. The first 4 channels are mainly designed for professional users. The Public channel provides China-VO selected resources and services for the public. It is the only one system in the world to provide resources and services to both professional and amateur astronomers, and cover the whole life cycle of astronomical Big Data.

Observation time application for astronomical telescopes is serving in the Observation channel. Features such as proposal submission, peer-review, time allocation, are provided. For special roles, such as telescope manager and PIs, additional functions are provided to them to manage proposals, time allocation committee and observation time, or to manage their observation data. Proposal submission sub-system was re-designed in 2015 to provide flexible management features for telescope managers. Configuration parameters for back-end instrument, proposal submission, peer-review, and observation time allocation can be changed by telescope managers easily.

Data channel is the portal for data exploration to access to both public data and datasets still in PI period. Tens of datasets hosted at the platform can be queried and used to carry out cross identified through a uniform interface [4]. LAMOST DR1 [5], the largest astronomical spectrum dataset with 2.2M spectra was open to the public in March 2015 through the system.

Well known and frequently used tools, packages, services, and source codes are collected into the Tools channel. Information and access links for more than one hundred of tools are provided, which are grouped into 7 categories, including data processing, data visualization, plotting, scripts, libraries, toolkits and VO tools.

Virtual Machines with pre-installed data processing and analysis software environments are available for users in Cloud channel. MyVOSpace and China-VO Paper Data repository are also serviced here. China-VO Paper Data repository provides long-term storage and open access service for user's paper data, which includes tables, figures, pictures, movies, source codes, models and software packages mentioned in his scientific papers. A permanent user specified URL will be provided for each item.

In 2015, China-VO prepared a popular astronomical data processing and analysis environment, i.e. MADARA. MADARA (Acquisition, Reduction and Analysis of Multi-wavelength Astronomical Data) is a Cloud Computing based teaching and research environment for astronomical lectures and graduate students. Common software packages, for example IRAF, DS9, CASA, HEASOFT, SSW, IDL, Python, can be used to process and to analyse multi-wavelength observation data. A virtual machine instance with these packages can be initiated and is ready for using in only few minutes from any selected Cloud node located in different cities.

Public channel is one modular especially developed for the public and amateur astronomers, including several VO featured applications and services. Video streams provide live images taken from video cameras at different observatories, for example the live image from Chinese Antarctic Zhongshan station. Gallery is a collection of beautiful pictures taken by AstroCloud users and amateur astronomers. Special is a collection of China-VO hosted services, for example Astronomical Dictionary, Chinese Starry Night and WWT Beijing Community.

On July 29, 2015, Popular Supernova Project (PSP) was lunched. It is the first astronomical citizen science project in China as a joint venture between China-VO and Xingming Amateur Astronomical Observatory. In the morning of Sep. 12, a supernova candidate was discovered by a 10-year old pupil. Inspired by the news, number of the registered users of AstroCloud platform raised violently up to 105K in the following several days. By the end of September 2016, 11 supernova and nova candidates have been discovered by public users, and among these 7 have been confirmed by professional observations.

A user dashboard is designed to give a user fast access points of frequently used resources and services, and summary information of the system and the user.

In additional to the above user facing channels, several crucial functions are provided by the backend platform [6]. Two examples are given here.

CSTNET passport: the combination of an email address and a password provided by China Science & Technology Network (CSTNET) that you use to sign in to supported services. If you don't have a CSTNET passport, you can sign up for free at any time.

Usage Statistics: give out important statistic data about the platform interested by users and administrators. For example, general weblog results and platform running status include online users, registered users, login in numbers, submitted proposals for telescopes, archived astronomical observation datasets and their latest progress, number of virtual machine instances at each Cloud computing node etc.

## V  CURRENT STATUS AND FUTURE PLANS

The China-VO AstroCloud was lunched on May 15, 2014. By the end of September 2016, the number of registered users has been up to 17.6 thousand. Latest usage statistic results are listed in table 1.

Through dedicatedly designed database and metadata management, the AstroCloud is acting as a whole lifecycle management platform for astronomical observation data (Fig. 8). Scientific outputs of a whole cycle of an astronomical observation, including proposal, observation log, raw data and data products are inter-connected and managed. Additionally, software tools and computing abilities are provided to process and analyze these data. A thin or even a mobile client is enough to do science on the platform.

In 2015, the five astronomical observatories in China were joined together to form Center of Astronomical Mega-Science, Chinese Academy of Sciences (CAMS), which will build Astronomical Technic Service Platform based on the China-VO AstroCloud platform. With the support of CAMS, AstroCloud will step into operation stage healthily. The role of AstroCloud will be enhanced from a whole life-cycle management platform for astronomical data to astronomical research planning platform for Chinese astronomy community.

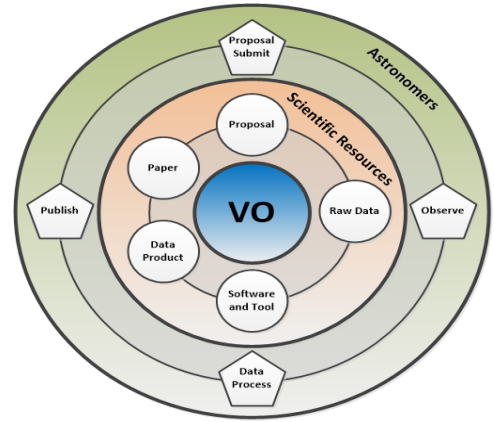

Figure 8.    Whole lifecycle management platform provided by the AstroCloud (VO)

TABLE I    . CURRENT USAGE STATUS OF THE ASTROCLOUD

| Feature | Status |
| --- | --- |
| Num. of Cloud nodes | 6 (Beijing, Nanjing, Shanghai, Kunming, Lijiang, Urumqi) |
| Num. of registered users | 17600+ |
| Num. of VM templates | 6 |
| Num. of user created VMs | 400 |
| Num. of telescopes for proposal submission | 4 |
| Num. of domestic observation datasets | 20+ |
| Num. of international datasets | 15000+ |
| Num. of discovered supernovas | 6 |
| Num. of discovered nova | 1 |

## ACKNOWLEDGMENT

Chinese Virtual Observatory (China-VO) is a jointed efforts of National Astronomical Observatories, Purple Mountain Observatory, Shanghai Astronomical Observatory, Yunnan Astronomical Observatories and Xinjiang Astronomical Observatory. Data resources are supported by Chinese Astronomical Data Center.